%% file: arXiv-Ver-3.tex
\documentclass[twocolumn,aps,prd,showpacs,nofootinbib,superscriptaddress,floatfix,showkeys]{revtex4-2}

\usepackage{amsfonts}
\usepackage{amsmath}
\usepackage{amssymb} 
\usepackage{bm} 
\usepackage{color}
\usepackage{enumerate}
\usepackage{graphicx}  
\usepackage{dcolumn}
\usepackage{hyperref}
\usepackage{lineno}

\hypersetup{
	colorlinks=true,       
	linkcolor=blue,          
	citecolor=magenta,        
	filecolor=cyan,      
	urlcolor=magenta           
}

\labelformat{section}{#1} 
\labelformat{subsection}{Section #1} 
\labelformat{figure}{Fig.~#1} 
\labelformat{subfigure}{Fig.~\thefigure#1} 
\labelformat{table}{Tab.~#1} 
\begin{document}
\title{Constraining circular polarization of high-frequency gravitational waves with CMB } 
\author{Ashu Kushwaha} 
\email{ashuk@iisc.ac.in}
\author{Rajeev Kumar Jain}
\email{rkjain@iisc.ac.in}
\affiliation{Department of Physics, Indian Institute of Science, C. V. Raman Road, Bangalore 560012, India}
%
	%
	%
    \date{\today}
\begin{abstract}
Circular polarization in the cosmic microwave background (CMB) offers a promising probe of the parity-violating physics of the early universe. In this paper, we propose a novel method to constrain the primordial circular polarization of high-frequency gravitational waves (GW) in the GHz range. An efficient conversion of gravitons to photons in a transverse cosmological magnetic field at the epoch of last scattering can generate excess chiral photons if the GW background is chiral in nature. This excess radiation distorts the CMB thermal black-body spectrum, which can be estimated by measuring the V-Stokes parameter in the CMB polarization. Using current upper limits on the angular power spectrum of circular polarization $C_l^{VV}$ from the CLASS, MIPOL, and SPIDER experiments, we obtain the most stringent constraints on the characteristic strain and circular polarization of the isotropic background of stochastic GWs at ${40\,\rm GHz}$ and ${150\,\rm GHz}$, respectively. Our work, therefore, provides an interesting possibility to constrain the circular polarization of high-frequency GWs using the V-mode polarization measurements of CMB. 

\end{abstract}
\maketitle

%
\textbf{Introduction.}
The cosmic microwave background (CMB) has played a pivotal role in advancing our understanding of the Universe over the past several decades. High-precision measurements from various CMB experiments have led to the establishment of the widely accepted $\Lambda$ Cold Dark Matter ($\Lambda$CDM) model. Anisotropies in the CMB’s intensity (temperature) and polarization (both linear and circular) provide crucial insights into the Universe’s origin, evolution, composition, and underlying physical mechanisms~\cite{Kosowsky:1994cy,Hu:1997hv,Kamionkowski:2015yta,Book-Durrer-2020,Komatsu:2022nvu}.
For instance, the detection of B-mode polarization is a powerful probe of primordial gravitational waves (GW), offering a means to constrain the energy scale of inflation. Moreover, a nonzero cross-correlation between E- and B-mode polarizations would signal parity-violating processes in the early Universe~\cite{Kosowsky:1994cy,Hu:1997hv,Kamionkowski:2015yta,Komatsu:2022nvu}. Since Thomson scattering at the last scattering surface generates only linear polarization (characterized by the Stokes $Q$ and $U$ parameters), the primordial circular polarization, described by the Stokes 
$V$ parameter is expected to vanish.
However, circular polarization can emerge through various mechanisms in the early Universe, such as photon-photon scattering~\cite{Sadegh:2017rnr,2019-Inomata.Kamionkowski-PRL,Hoseinpour:2020hic}, photon-graviton scattering~\cite{Bartolo:2018igk}, interactions in a magnetized plasma~\cite{Giovannini-PhysRevD.80.123013,Giovannini:2010ar}, Faraday conversion~\cite{Finelli:2008jv,De:2014qza,Book-Durrer-2020}, and parity-violating physics~\cite{Alexander:2017bxe,Alexander:2018iwy,Alexander:2019sqb}. Detecting a nonzero primordial circular polarization signal would provide significant constraints on beyond-standard-model (BSM) physics, particularly those involving parity violation in the early Universe.
Recently, notable constraints on the angular power spectrum of primordial circular polarization,
$C_l^{VV}$, have been reported by several CMB experiments, including the Milano Polarimeter (MIPOL) \cite{Mainini:2013mja}, SPIDER \cite{SPIDER:2017kwu}, and the Cosmology Large Angular Scale Surveyor (CLASS) \cite{Padilla:2019dhz-CLASS,Eimer_2024-CLASS} which represent an important step towards constraining new physics through CMB polarization measurements.

Due to their weakly interacting nature, GWs decouple immediately after their production. Thus, constraining the properties of the stochastic GW background (SGWB) provides a direct probe of the underlying generation mechanisms in the early Universe ~\cite{1999-Maggiore-PhyRept,2004-Bisnovatyi.Kogan-CQG,2009-Sathyaprakash.Schutz-LivRevRel,Book-Maggiore-Vol2,Dong:2015yjs,2018-Caprini.Figueroa-CQG,2021-Domenech-Universe-GW-Review,Mandal:2025xuc}. 
However, astrophysical sources of high-frequency GWs (HFGWs) in the MHz to GHz range, such as exotic compact objects, primordial black holes, and the early Universe physics, remain unknown, and their experimental detection is still in its early stages, with only a few detectors currently operational~\cite{2006-Cruise-CQG,2008-Nishizawa.etal-PRD,2012-Cruise-CQG,2017-Chou.etal-PRD,2020-Aggarwal.etal-LivingRevRel,Domcke:2022rgu,Bringmann:2023gba,Aggarwal:2025noe}.
Recent advancements have proposed various indirect detection methods, primarily utilizing radio observations, to place constraints on linearly polarized HFGWs~\cite{2021-Domcke.Garcia-Cely-PRL,2022-Kushwaha.etal-MNRAS,2023-Kushwaha.Sunil.Shanki-IJMPD,Ito:2023nkq,Ito:2023fcr,2024-He.Sharma.etal-JCAP}. However, these studies have largely overlooked the circular polarization of the SGWB. A nonzero circular polarization, indicative of a preferred handedness, would provide compelling evidence of parity-violating processes in the early Universe, offering an essential probe of fundamental physics.
Parity-violating sources induce the asymmetrical evolution of the two helicity modes of propagating waves, which results in the suppression of one helicity mode while amplifying the other, a phenomenon known as amplitude birefringence~\cite{1999-Lue.Wang.Kamionkowski-PRL,2003-Jackiw.Pi-PRD,2009-Alexander.Yunes-PhyRept,2010-Gluscevic.Kamionkowski-PRD,2011-Durrer.Hollenstein.Jain-JCAP,2012-Byrnes.etal-JCAP,Inomata:2018vbu,2020-Kushwaha.Shankaranarayanan-PRD,2021-Okano.Fujita-JCAP,Komatsu:2022nvu,Fu:2024ipa}. Consequently, the generated waves become circularly polarized, a feature extensively studied as a probe of BSM physics~\cite{Book-Durrer-2020,Komatsu:2022nvu}.
A recent claim of observed parity violation in the four-point function of BOSS galaxy clustering data~\cite{Hou:2022wfj,Adari:2024vkf,Inomata:2024ald} strongly motivates the development of theoretical frameworks to directly or indirectly constrain parity-violating signatures in the matter and gravity sectors. 

In this work, we present a novel methodology for constraining the circular polarization of the SGWB using recent measurements of the angular power spectrum of the Stokes V-mode polarization from various CMB experiments.
Our formalism relies on the conversion of gravitons to photons in a cosmological magnetic field and provides model-independent insights into parity-violating physics in the gravity sector of the early Universe. To the best of our knowledge, no previous studies have explored or established any constraints on the circular polarization of the SGWB at high frequencies, which highlights the significance of our work. 

\textbf{Graviton-photon conversion.}
We consider a scenario wherein the incoming GWs convert to electromagnetic (EM) waves in the presence of a transverse magnetic field, a phenomenon known as the \textit{Gertsenshtein-Zel’dovich effect}~\cite{1962-Gertsenshtein-JETP,1974-Zeldovich-SJETP,Chen:1994ch,Bastianelli:2004zp,Bastianelli:2007jv,2013-Chen.Suyama-PRD,2012-Dolgov.Ejlli-JCAP,2013-Dolgov.Ejlli-PRD,2020-Fujita.Kamada.Nakai-PRD,2021-Domcke.Garcia-Cely-PRL,2022-Kushwaha.etal-MNRAS,2023-Kushwaha.Sunil.Shanki-IJMPD,2023-Palessandro.Rothman-PDU,2023-Dolgov.etal-Universe,2024-He.Sharma.etal-JCAP}.
For waves with a frequency $\omega$ greater than the Hubble parameter $H = \dot{a}/a$, we can ignore the expansion of the Universe. Thus, the graviton-photon oscillation system in Minkowski spacetime is described by coupled Einstein-Maxwell equations (hereafter, we often refer to our work~\cite{2024-Kushwaha.Jain-PRD} for notations and detailed calculations)
\begin{align}\label{gw-em-main}
	\!\!  \left( \Box - \omega_{Pl}^2 \right) A_{\sigma}  &= - \sigma i\bar{B} \partial_z h_{-\sigma} , ~~ \Box h_{\sigma} = \frac{2 \sigma i}{M_P^2} \bar{B} \partial_z A_{-\sigma} 
\end{align}
where $\sigma = \pm$ (or $R/L$) refer to the helicity of propagating GWs (tensor fluctuations, $h_{\sigma} = \sqrt{2} \, \tilde{h}_{\sigma}/M_P$) and EM waves\footnote{We define $A_{\pm} = (A_x \mp i A_y)/\sqrt{2}$ and $h_{L/R} = (h_+ \mp ih_{\times})/\sqrt{2}$~\cite{2024-Kushwaha.Jain-PRD}.} (vector field fluctuations, $A_{\sigma}$), and $\bar{B}$ is the transverse background (cosmological) magnetic field (see Refs.~\cite{2001-Grasso.etal-PhyRep,2002-Widrow-Rev.Mod.Phys.,2004-Giovannini-IJMPD,2007-Martin.Yokoyama-JCAP,2007-Barrow.etal-PhyRep,2011-Kandus.Kunze.Tsagas-PhysRept,2013-Durrer.Neronov-Arxiv,2013-Ferreira.Jain.Sloth-JCAP,Ferreira:2014hma,2016-Subramanian-Arxiv,Tripathy:2021sfb,Tripathy:2022iev}).
Since we consider the mechanism to occur during the decoupling epoch, the interaction of photons with the plasma changes the dispersion relation between photon momentum ($k$) and its frequency ($\omega$) as $ \omega_{Pl}^2 = \omega^2 - k^2 \neq 0$ where the plasma frequency, $\omega_{Pl}^2 = e^2 n_e/m_e$. Furthermore, assuming a general dispersion relation of the form $k=n\omega$ with refractive index given by $n = \sqrt{1-\frac{ \omega_{Pl}^2}{\omega^2}}$ satisfying $|1-n| \ll 1$, we can approximate $\omega-i\partial_z = \omega + k \simeq 2\omega$ where  $-i\partial_z = k$. 

Note that, Eq.~\eqref{gw-em-main} describes the conversion of gravitons to photons in two channels: $(\tilde{h}_L \leftrightarrow A_+)$ and $(\tilde{h}_R \leftrightarrow A_-)$. This can also be interpreted as the transfer/oscillation of chirality between the GW and EM sectors, see Ref.~\cite{2024-Kushwaha.Jain-PRD}.
For waves propagating along the $z-$direction with a specific frequency of the modes, i.e., $ h_{\sigma} (z,t) = h_{\sigma} (z) e^{-i\omega t}$ and  $A_{\sigma} (z,t) = A_{\sigma} (z) e^{-i\omega t}$, the graviton-photon system~\eqref{gw-em-main} for the channel $(\tilde{h}_L \leftrightarrow A_+)$ is described by linearized differential equations, which in the matrix form is given by~\cite{2024-Kushwaha.Jain-PRD}
\begin{align}\label{oscillation-eq-matrix}
	\left[ (\omega + i\partial_z) \mathbb{I} + \mathcal{M} \right]   \hat{\Psi} (z)  \simeq 0 ~,
\end{align}
where $\mathbb{I}$ is the identity matrix, $ \hat{\Psi} (z) = \begin{bmatrix}
	A_+ (z) \\ \tilde{h}_L (z)
\end{bmatrix} $ is the column vector field, and $\mathcal{M}$ is given by the matrix
\begin{align}\label{mixing-matrix-def}
  \mathcal{M} = \begin{bmatrix}
        -\omega_{Pl}^2 /2\omega & -\bar{B}/\sqrt{2} M_P \\
        -\bar{B}/\sqrt{2} M_P & 0
    \end{bmatrix}
\end{align}
with eigenvalues $\lambda_{1,2} = - \frac{1}{2} \left[ \frac{\omega_{Pl}^2 }{2\omega} \pm \sqrt{\left( \frac{\omega_{Pl}^2}{2\omega} \right)^2 + \frac{2 {\bar{B}}^2}{M_P^2}} \right]$.
Similarly, we can obtain the matrix equation for the other channel (i.e., $\tilde{h}_R \leftrightarrow A_-$). Furthermore,
assuming that there are only gravitons at the beginning and \emph{no} photons as the initial condition allows us to derive a simple relation between both channels $ \langle \tilde{h}_R (0) | A_- (z) \rangle = \frac{1}{\delta_g}  \langle \tilde{h}_L (0) | A_+ (z) \rangle$, where $|\tilde{h}_L | = \sqrt{\delta_g} |\tilde{h}_R|$ and $\delta_g$ are the parameters that determine the imbalance between left- and right-circular modes of GWs and hence their chirality~\cite{2024-Kushwaha.Jain-PRD}. The probability of conversion of gravitons to photons (for both channels) can be calculated as (see the appendix~\ref{appendix} for more details)
\begin{align}\label{conv-probability-2a}
	\mathcal{P}^+_{g \rightarrow \gamma} (\ell) = \delta_g^2 \mathcal{P}^-_{g \rightarrow \gamma} (\ell) =   \frac{\bar{B}^2 \ell^2_{osc}}{2 M_P^2} ~ \sin^2\left[ \frac{\ell}{\ell_{osc}}\right] ~~,
\end{align}
where $|z| = \ell$ is the typical length scale over which the conversion of GWs to EM waves is estimated and $\ell_{osc}^{-1} = \frac{1}{2} \sqrt{ \omega^2 (1-n)^2 + \frac{2 {\bar{B}}^2}{M_P^2} } $. 
Since the cosmological magnetic fields are not perfectly homogeneous at all length scales, the efficient conversion takes place only in homogeneous regions for which $\ell_{osc} \ll \ell$, hence the sinusoidal term can be averaged to $\sin^2 \left( \ell/\ell_{osc} \right) \sim \frac{1}{2}$. Note that Eq.\eqref{conv-probability-2a}, is suitable for the propagation distance up to the coherence length of the magnetic field.  
For instance, if the propagation distance $D$ of GWs is larger than the coherence length of the magnetic fields, 
the magnetic field would be inhomogeneous over the distance $D$. In that case, we obtain cumulative effects from all domains of a typical length scale $\ell$ with a uniform magnetic field. The number of these homogeneous domains is given as $\Delta N \simeq D/\ell$ with the condition $\mathcal{P}^{\pm}(\ell) \, \Delta N \ll 1$~\cite{2021-Domcke.Garcia-Cely-PRL}. Thus, the conversion probability can be obtained as 
\begin{align}\label{conv-probability-tot-lambda-B}
\langle	\mathcal{P}^+_{g \rightarrow \gamma} (\ell)  \rangle = \delta_g^2 \langle	\mathcal{P}^-_{g \rightarrow \gamma} (\ell)  \rangle  \simeq   \left(  \frac{\bar{B} \,  \ell_{osc}}{2 M_P} \right)^2 \left( \frac{D}{\ell} \right) ~~.
\end{align}
Furthermore, we can calculate the average probability per unit time  (i.e., average conversion rate) for GW to propagate over a distance $D$ in time $\Delta t$ as \begin{align}
\langle \Gamma^+_{g \rightarrow \gamma } \rangle   \equiv \frac{ \langle 	\mathcal{P}^+_{g \rightarrow \gamma} (\ell) \rangle }{D} =   \frac{1}{\ell}  \left(  \frac{\bar{B}\, \ell_{osc}}{2 M_P} \right)^2 = \delta_g^2 \langle \Gamma^-_{g \rightarrow \gamma } \rangle 
\end{align}
The total conversion probability can be obtained by integrating over the line of sight of propagation, which in the expanding universe can be given by

\begin{align}\label{tot-con-prob}
    \mathbb{P}^{\pm} \equiv \int_{l.o.s} \langle \Gamma^{\pm}_{\gamma \rightarrow g} \rangle  dt =  \int_0^{z_i} \frac{\langle \Gamma^{\pm}_{\gamma \rightarrow g} \rangle }{(1+z) H} dz
\end{align}
where $H\,dt = dz/(1+z)$, hereafter, the quantity $z$ refers to the cosmological redshift and should not be confused with the Cartesian coordinate.

Since we focus on the decoupling epoch (i.e., $z_{\rm dec} \simeq 1100$), we have the initial redshift, $z_i \leq z_{\rm dec}$ and the Hubble parameter during this epoch (matter-domination) is $H = H_{\rm dec} \left( \frac{T}{T_{\rm dec}} \right)^{3/2}$, where $T_{\rm dec} = 0.26 \, {\rm eV}$ is the temperature of the universe at decoupling.
Note that during this epoch the refractive index of the medium for the high-frequency CMB photons is determined by the electron density, neutral hydrogen, and helium.
The electron number density during this epoch is given as $n_e (z)  = n_{b0} (1+z)^3 \, X_e (z)$, where $n_{b0} = 0.251 \, {m}^{-3} = 1.928\times 10^{-48} \, {\rm GeV}^3$ is the baryon number density today~\cite{2021-Domcke.Garcia-Cely-PRL}. Using these estimates we obtain the plasma frequency as $\omega_{Pl} (z) \simeq 27.67 \, {\rm Hz} \, (1+z)^{3/2} X_e^{1/2} (z)$. 
We use the values of ionization fraction $X_e (z)$ from Ref.\cite{Kunze:2015noa}, which at $z=0, 1100$ are $1, 0.154$ respectively. 
Since the CMB photon frequency today is in the ${\rm GHz}$ range and the plasma frequency at the present epoch is $\omega_{Pl,0}\sim \,  27.67 \, {\rm Hz}$, we can make the approximation for the refractive index as $1 - n(z) \simeq \frac{ \omega_{Pl}^2}{2 \omega^2}$.
The frequency of CMB photons redshifts as $\omega \propto 1/a \sim (1+z)$ 
and using the relation for the plasma frequency $\omega_{Pl}^2 = \omega_{Pl,0}^2  (1+z)^3 X_e (z) $, we obtain $ 1 - n(z) \simeq (1+z) \, X_e (z) \, \frac{ \omega_{Pl,0}^2}{2 \omega_0^2}$.
Furthermore, one can verify that even for a nano-Gauss present-day magnetic field strength, $\bar{B}_0 = 1 \, {\rm nG}$ and $\omega_0 = 1 \, {\rm GHz}$, the term $2\bar{B}^2/M_{Pl}^2$ is much smaller compared to $ \left( \frac{\omega_{Pl}^2}{2\omega} \right)^2$. Therefore, we can approximate $	\ell_{osc}^{-1}  \simeq (1+z)^2 X_e (z) \frac{ \omega_{Pl,0}^2}{4 \omega_0}$.
Using the decaying behaviour of magnetic fields, $\bar{B} (z) = \bar{B}_0 (1+z)^2$ and $\ell = \ell_0/(1+z)$, the total conversion probability in Eq.\eqref{tot-con-prob} can be obtained as
\begin{align}\label{tot-con-prob-plus-minus}
    \mathbb{P}^{+} = \delta_g^2 \mathbb{P}^{-} = & \;4.04\times 10^{-19} \, h^{-1} \left( \frac{{\rm Mpc}}{\ell_0} \right) 
    \nonumber\\
    & \times \left(  \frac{\bar{B}_0 }{1 {\rm nG}} \right)^2  \left(\frac{\omega_0}{ T_0} \right)^{2} \left( \frac{\mathcal{I} (z_i)}{10^6} \right),
\end{align}
where $H_0 = 2.13\times 10^{-42} h \, {\rm GeV}$ is the Hubble constant at the present epoch and the redshift-dependent integration term $ \mathcal{I} (z_i) = \int_0^{z_i} (1+z)^{-3/2} X_e^{-2} (z) dz \sim 6.31\times 10^6$ at $z_i = 1100$.

Before decoupling the CMB photons were in thermal equilibrium and are described by the Bose-Einstein distribution function $f_{\gamma}^{\rm eq} = 1/(e^{\omega/T}-1) $. Due to graviton-photon conversion, some gravitons will be converted to photons, which would affect the spectrum of CMB photons, causing the distortions of the thermal blackbody spectrum. We can quantify this distortion by estimating the difference $\Delta f_{\gamma} = f_{\gamma} - f^{\rm eq}_{\gamma}$, where $f_{\gamma}$ is the total photon distribution. Assuming before the onset of graviton-photon conversion $\Delta f_{\gamma} = 0$, which essentially means the distortion is only caused due to this mechanism. 
Next, we can define the energy spectrum of the gravitons (or the SGWB) which is characterized by the quantity $\Omega_{GW}$, the energy density of gravitons per logarithmic frequency bin at present epoch~\cite{2018-Caprini.Figueroa-CQG,2021-Domcke.Garcia-Cely-PRL}
\begin{align}\label{gwb-definition}
	\Omega_{GW} \equiv \frac{1}{\rho_c} \frac{d \; \rho_{GW} (\omega,T)}{d \ln \omega}, \rho_{GW} (T) \equiv \int \frac{d\ln\omega ~ }{\pi^2} \omega^4 f_g (\omega,T)
\end{align}
where $\rho_c = \frac{3H^2}{8\pi G}$ is the critical density of the Universe, and its value at present epoch is $\rho_c^0 = \frac{3H_0^2}{8\pi G} = 8.098 h^2\times 10^{-47}\, {\rm GeV}^4$, and $H_0 = 100 \, h \,{\rm km \, s^{-1} Mpc^{-1}}$ is the Hubble parameter at the present epoch~\cite{Book-Dodelson-2020,Book-Baumann-2022}. 
From the above equation and using $\Omega_{\gamma} = \frac{\rho_{\gamma}}{\rho_c} = \frac{\pi^2 T^4}{15 \, \rho_c}$ where $\rho_{\gamma}$ is the energy density of photons, we obtain the distribution function for gravitons as $f_g (\omega,T) = \frac{\pi^4}{15} \left( \frac{T}{\omega} \right)^4 \frac{\Omega_{GW}}{\Omega_{\gamma}}$.
The evolution of the distribution functions is described by the Boltzmann equation~\cite{2021-Domcke.Garcia-Cely-PRL,2024-He.Sharma.etal-JCAP,2024-Kushwaha.Jain-PRD}
\begin{equation}\label{boltzmannEq-T}
	- H \left( T \partial_T + \omega \partial_{\omega}  \right) f_{\gamma,g} (T,\omega) = \pm \langle \Gamma_{\gamma \rightarrow g} \rangle (f_g - f_{\gamma} )
\end{equation}
where $ \langle \Gamma_{\gamma \rightarrow g} \rangle $ is the average probability per unit time\footnote{Here we suppress the notation for left-right photon modes to avoid confusion. We shall recover them after this straightforward algebra, which is the same for both modes}. 
Following Ref. \cite{2021-Domcke.Garcia-Cely-PRL,2024-He.Sharma.etal-JCAP}, the solution of Eq.\eqref{boltzmannEq-T}, is given by $\Delta f_{\gamma}^{\pm} (\omega_0 , T_0) = \left[ f_g (\omega_i , T_i) - f^{\rm eq}_{\gamma} \right] \mathbb{P}^{\pm}$, which can be written as
\begin{align}\label{sol-photon-distortion-fraction}
    \frac{\Delta f_{\gamma}^{\pm} (\omega_0 , T_0) }{f^{\rm eq}_{\gamma} }  = \left[ \frac{\pi^4}{15} \left( \frac{T_0}{\omega_0} \right)^4 \frac{\Omega_{GW}}{\Omega_{\gamma}} \left( e^{\omega_0/T_0} -1 \right) - 1 \right] \mathbb{P}^{\pm} 
\end{align}
where $\omega_0/T_0 = \omega/T$ and $\Omega_{GW}/ \Omega_{\gamma} = \Omega_{GW,0}/ \Omega_{\gamma,0}$.
As we know, the GWs would contribute to the total energy density of the Universe in the form of relativistic degrees of freedom (i.e., radiation), which are well constrained by Big-Bang Nucleosynthesis (BBN) and CMB bounds on the effective number of degrees of freedom~\cite{2018-Caprini.Figueroa-CQG,2021-Domcke.Garcia-Cely-PRL} $ \frac{\Omega_{GW}}{\Omega_{\gamma}} \lesssim \frac{7}{8} \left( \frac{4}{11}\right)^{4/3} \Delta N_{\rm eff} $, %
where $\Delta N_{\rm eff} \lesssim 0.1$~\cite{RevModPhys.88.015004,Planck:2018vyg,2021-Domcke.Garcia-Cely-PRL}.

The photons due to the conversion mechanism determined by Eq.~\eqref{sol-photon-distortion-fraction} would contribute to the excess of background EM radiation which can be quantified in terms of the intensity ($I_{\omega}$) of photons in the frequency range $\omega$ to $\omega + d\omega$ as $	\Delta I_{\omega}  (\omega,T) =  \frac{\omega^3}{4\pi^2 } \Delta f_{\gamma} (\omega,T) $~\cite{Book-Baumann-2022}.
This provides the relation between the change in total intensity $\Delta I = \Delta I_+ + \Delta I_-$ and the change in the spectrum $\Delta f_{\gamma} = \Delta f_{\gamma}^+ + \Delta f_{\gamma}^- $ at a fixed given frequency $\omega$.
The CMB experiments usually measure the spectral radiation intensity $I_{\omega} $, which is the energy flux per unit area per unit frequency. 
The fractional change in the intensity of excess radiation with respect to the CMB photons in thermal equilibrium (blackbody spectrum) can be determined by the relation $\frac{\Delta I^{\pm}}{I^{\rm eq}_0} = \left( \frac{\omega_0 }{\omega_0^{\rm eq}}\right)^3 \frac{\Delta f_{\gamma}^{\pm}}{f^{\rm eq}_{\gamma}}$, where $I^{\rm eq}_0$ is the peak intensity of CMB photons for $f_{\gamma}^{\rm eq}$ at the present epoch, i.e., $T_{\gamma,0} \simeq 2.726 \, K \simeq 354 \, {\rm GHz}$ and $ \omega_0^{\rm eq} \simeq 2.82 \, T_{\gamma,0} \simeq 998 \, {\rm GHz}$~\cite{2024-Kushwaha.Jain-PRD}.
Thus, using Eq.\eqref{sol-photon-distortion-fraction}, we obtain the total fractional intensity of excess radiation as
\begin{align}\label{excess-intensity-final}
	\frac{\Delta I}{I^{\rm eq}_0} = & \left( \frac{\omega_0 }{\omega_0^{\rm eq}}\right)^3 \,  \left[ \frac{\pi^4}{15} \left( \frac{T_0}{\omega_0} \right)^4 \frac{\Omega_{GW}}{\Omega_{\gamma}} \left( e^{\omega_0/T_0} -1 \right) - 1 \right] \nonumber
    \\
    & \times \left[\frac{2 (1 + \Delta \chi_g^2)}{(1-\Delta \chi_g)^2}\right] \, \times 4.04\times 10^{-19} \, h^{-1} \left( \frac{{\rm Mpc}}{\ell_0} \right) 
    \nonumber \\
    & \times \left(  \frac{\bar{B}_0 }{1 {\rm nG}} \right)^2  \left(\frac{\omega_0}{ T_0} \right)^{2} \left( \frac{\mathcal{I} (z_i)}{10^6} \right) ~.
\end{align}
Note that for a given measurement of excess radiation ($\Delta I/I^{\rm eq}_0$) at the frequency $\omega_0$, the above relation can be used to provide the constraints on the parameter space of the amplitude of the polarized GW background ($\Omega_{GW}$) with net circular polarization $\Delta \chi_g$. 

\begin{figure*}[t!]
\centering
\includegraphics[height=2.8in,width=3.45in]{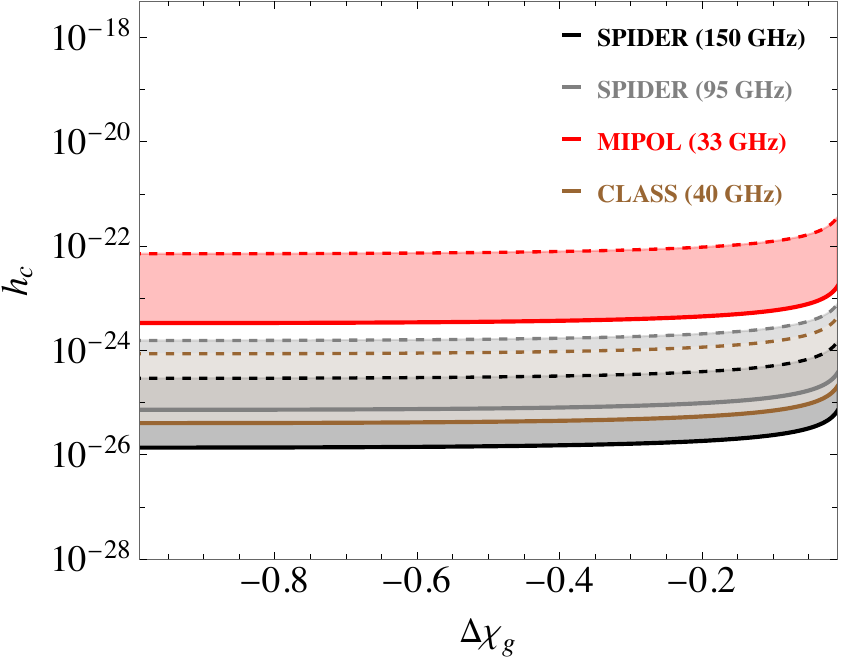} 
\hspace{.2cm}
\includegraphics[height=2.8in,width=3.45in]{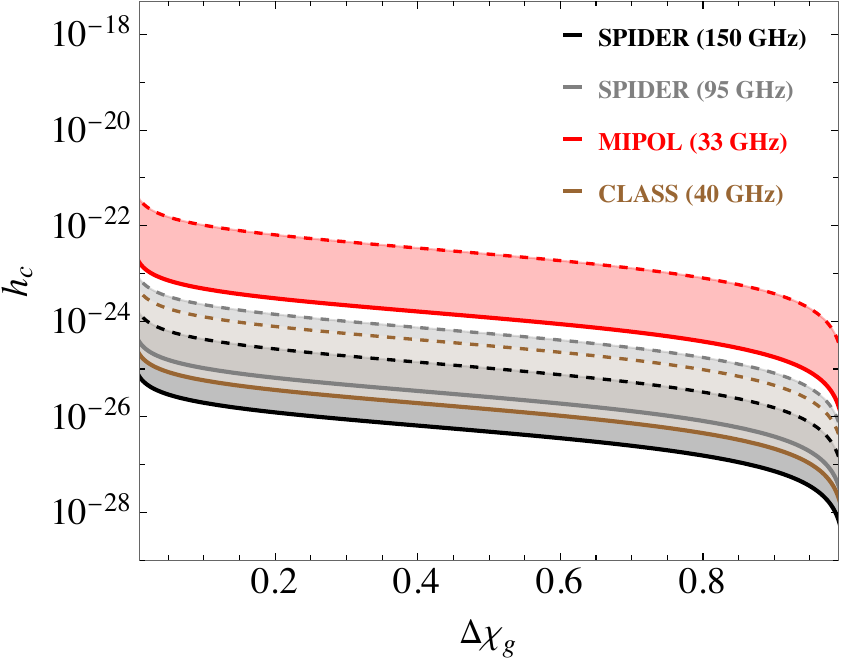} 
\caption{The various constraints (upper limits) on the characteristic strain $h_c$ of the polarized GW background for three CMB experiments SPIDER, MIPOL and CLASS at different frequencies and multipoles are plotted for a negative value of $\Delta \chi_g$ (on the left) and for positive $\Delta \chi_g$ (on the right). The shaded region shows the range of values for magnetic field strength $B_0$ spanning from $ 47 \, {\rm pG}$ (dashed line) to $ 1 \, {\rm nG}$ (solid lines) with a coherence length $\ell_0 = 1\, {\rm Mpc}$. While the constraints from MIPOL are the weakest, SPIDER and CLASS provide comparable constraints on $h_c$.}
\label{fig:constraint-plot}
\end{figure*}

The circular polarization of the background GWs and the produced photons is defined as $\Delta \chi_g = \frac{1-\delta_g}{1+\delta_g}$ and $\Delta \chi_{\gamma} =  \frac{1-\delta_A}{1+\delta_A}$, respectively (see, for instance, our earlier work~\cite{2024-Kushwaha.Jain-PRD}). Note that $\Delta \chi_{g,\gamma} = + 1 (-1)$ refers to the maximally helical right-handed (left-handed) waves. 
Using Eq.~\eqref{tot-con-prob-plus-minus} which gives the conversion probability of propagating GWs to photons i.e., $\mathbb{P}^{+} = \delta_g^2 \mathbb{P}^{-}$, we can derive the relation between the energy density carried by both the helicity modes of the produced photons as $\Omega_{\gamma}^{-} =  \frac{1}{\delta_g^2} \,\,  \Omega_{\gamma}^+$ which implies $ \delta_A = \frac{1}{\delta_g^2}$.
Note that, this simple quadratic dependence is one of the distinguishing features of the chirality transfer between the GW and EM sectors in the graviton-photon oscillation mechanism as discussed in our earlier work~\cite{2024-Kushwaha.Jain-PRD}. This asymmetric relation arises due to the preference over the dominant helicity mode of circularly polarized GWs (i.e., the normalization factor in Eq.\eqref{R-minus-L-plus-eqs}, which determines the helicity of the propagating GWs. See the appendix~\ref{appendix} for more details. Furthermore, it allows us to compute an interesting yet non-trivial relation between $\Delta \chi_g$ and $\Delta \chi_{\gamma}$ as
\begin{equation}\label{chirality-parameter-relation-main}
	\Delta \chi_{\gamma} =  -\frac{2 \Delta \chi_{g}}{1+\Delta \chi_{g}^2} ~~,
\end{equation}
where the chirality parameter of GWs is defined in the range $-1 < \Delta \chi_g < 0$ and $0 < \Delta \chi_g <1$~\cite{2024-Kushwaha.Jain-PRD}.
It is interesting to note that, $\Delta \chi_{\gamma} \to 0$ as $\Delta \chi_{g} \to 0$ but $\Delta \chi_{\gamma} \to \pm 1$ as $\Delta \chi_{g} \to \mp 1$.

\textbf{Constraints from CMB circular polarization measurements.}
The polarization of the radiation is determined by the Stokes parameters which are defined in terms of two orthogonal components of the electric field, which in flat FRW spacetime are given by
\begin{align}\label{stokes-para-def}
	I = \frac{1}{a^2} \left( |E_+|^2 + |E_-|^2 \right), ~
	V =  \frac{1}{a^2}  \left( |E_+|^2 -  |E_-|^2 \right)
\end{align}
where $a(t)$ is the scale factor, $I$ refers to the intensity of the polarized radiation flux (approximately equal to the average intensity of the radiation~\cite{Book-Baumann-2022}) and $V$ refers to the circular polarization. The Stokes parameters have units of intensity, however, because CMB is observed to be (nearly) perfect blackbody, the temperature can uniquely determine the intensity. Therefore, often the fluctuations (anisotropies) in the CMB experiments are measured in units of temperature~\cite{2002-Staggs.etal-proceeding}.

In Eq.\eqref{stokes-para-def}, the total intensity of the radiation (photons) is determined by $I = I_0 + \delta I_{\rm CMB} + \Delta I_{g\rightarrow \gamma}$, where $I_0$ is the peak intensity of the background CMB photons (corresponding to $T_{\gamma,0} = 2.726 \, K $), $\delta I_{\rm CMB}$ is the intensity corresponding to the distortion in the CMB spectrum due to various other mechanisms~\cite{Kosowsky:1994cy,Hu:1997hv,Komatsu:2022nvu}, and $\Delta I_{g\rightarrow \gamma}$ is the intensity of the excess photons due to graviton-photon conversion (this work). 
Similarly, we can write the circular polarization Stokes parameter $V = V_0 + \Delta V$, where $V_0 = 0$ as the circular polarization is not produced via Thomson scattering at the background level, and $\Delta V$ is the contribution due to graviton-photon conversion mechanism (this work). 
Assuming that the graviton-photon conversion mechanism is the only source of excess polarized radiation, we find the Stokes parameters as $ \Delta I = \Delta I_{g\rightarrow \gamma}  = \frac{| \Delta E_+|^2}{a^2} \left( 1 + \delta_A \right)$ and $\Delta V  = \frac{| \Delta E_+|^2}{a^2} \left( 1 - \delta_A \right)$, where $\delta_A = (|A_-|/|A_+|)^2$~\cite{2024-Kushwaha.Jain-PRD}. 
Furthermore, using the definition of the chirality parameter of the photons gives the following relation
\begin{align}\label{delta-V-1-relation}
	\Delta V = \Delta \chi_{\gamma} \cdot \Delta I ~~.
\end{align}
In the above equation, Stokes-V and I have intensity units. Following Refs.~\cite{Alexander:2017bxe,2002-Staggs.etal-proceeding}, we can define the V-mode in the temperature units where the ratio is $\frac{V}{ I} = \frac{ V_T}{ T}$. Since the CMB experiments measure the V-mode in temperature units, $V_T$ is our observable quantity. Using Eq.\eqref{chirality-parameter-relation-main} and Eq.\eqref{delta-V-1-relation}, the power spectrum of the V-mode can be obtained by the two-point correlation function as~\cite{Alexander:2017bxe}
\begin{align}
	\Bigg\langle \frac{\Delta V_T}{T_{\gamma,0}} \, \frac{\Delta V_T}{T_{\gamma,0}} \Bigg\rangle  = \Bigg\langle \left( \frac{2 \Delta \chi_{g}}{1+\Delta \chi_{g}^2} \right)^2 \, \frac{\Delta I}{I^{\rm eq}_0} \, \frac{\Delta I}{I^{\rm eq}_0}  \Bigg\rangle ~~.
\end{align}
Since the CMB experiments measure the quantity 
$\langle \Delta V_T \, \Delta V_T \rangle = \frac{l (l+1)}{2\pi} C_l^{VV} T_{\gamma,0}^2$, therefore, in the above equation, the LHS is fixed for a given CMB measurement. Thus, using Eq.\eqref{excess-intensity-final} for $\Delta I/I^{\rm eq}_0$ on the RHS of the above expression at a given measurement frequency would constrain the parameter space 
the polarization $\Delta \chi_g$ and the characteristic strain $h_c = \sqrt{\frac{3H_0^2}{2\pi^2} \left(\frac{2\pi}{\omega_0}\right)^2 \Omega_{GW,0}}$. We use the most updated constraints on the V-mode angular power spectrum $\frac{l (l+1)}{2\pi} C_l^{VV} T_{\gamma,0}^2$ from three different CMB experiments to obtain an upper limit on the ($h_c - \Delta \chi_g$) parameter space.

Recently, a ground-based mission CLASS has reported the most stringent constraints at multipole range $5 < l < 125$ at $\nu_0 = 40 \, {\rm GHz}$ frequency ($\nu_0 = \omega_0/2\pi$) 
as $ \frac{l (l+1)}{2\pi} C_l^{VV} T_{\gamma,0}^2 < 0.023 \, {\mu K^2}$~\cite{Eimer_2024-CLASS}.
The balloon-borne experiment SPIDER~\cite{SPIDER:2017kwu} constraints at higher multipole (in the range $33 < l < 307$) at frequencies $95\, {\rm GHz}$ and $150\,{\rm GHz}$ are $\leq 783 \, {\mu K^2}$ and $\leq 141 \, {\mu K^2}$ respectively.
A much weaker constraint at $33 \, {\rm GHz}$ (at lower multipole) is also reported from MIPOL constraints as $ \leq 2\times 10^5 \, {\mu K^2}$~\cite{Mainini:2013mja}.
We use these constraints mentioned above to obtain the upper limits on the circular polarization of the SGWB at higher-frequency ranges, which are displayed in \ref{fig:constraint-plot}.
The shaded region shows the range of values for magnetic field strength $B_0$ from $ 47 \, {\rm pG}$ to $ 1 \, {\rm nG}$ with a typical coherence length $\ell_0 = 1\, {\rm Mpc}$, which are derived from the CMB constraints~\cite{Planck:2015zrl}. We can see that the constraints from MIPOL are the weakest, the SPIDER and CLASS provide stringent constraints at $150 \, {\rm GHz}$ and $40 \, {\rm GHz}$ on the amplitude of the characteristic strain $h_c $, as a function of $\Delta \chi_g$.

\textbf{Discussion.}
The detection of HFGWs presents an exciting opportunity to probe BSM physics, particularly due to their potential origin from unknown astrophysical or cosmological sources. However, direct detection of HFGWs, especially in the GHz range, is still in the preliminary stages, with current technologies requiring significant advancements that may take decades to become operational. 
In this context, constraining the circularly polarised background of HFGWs provides a crucial opportunity to probe parity-violating processes in the early Universe, offering insights into a class of BSM physics.
Given the current technological limitations, the development of new indirect methods to probe circularly polarized HFGWs using existing observations becomes essential. This approach is a viable alternative to direct detection while serving as an independent and complementary means to refine the parameter space allowed for such new physics.

In this work, we have emphasized that CMB experiments designed to measure circular polarization (V-mode) can serve as powerful indirect probes of HFGWs.
Specifically, we have demonstrated how circularly polarized GWs can generate an excess circularly polarised signal (radiation) in the CMB through the graviton-to-photon conversion mechanism, which can be measured as the V-mode. 
By utilizing recent upper limits on the V-mode angular power spectrum from MIPOL, SPIDER, and CLASS, we placed the most stringent constraints to date on the parameter space of characteristic strain and chirality parameter at $40\, {\rm GHz}$ and $150 \, {\rm GHz}$.
Therefore, these CMB polarization experiments offer valuable insights into the physics of the early Universe and BSM physics, making them crucial tools in the search for new fundamental interactions.
Despite significant progress in polarization studies, circular polarization has received relatively little attention, as most experiments have focused predominantly on linear polarization (E and B modes). We argue that shifting the scientific focus of future missions towards V-mode measurements could provide valuable insights -- not only into the physics of the early Universe but also into BSM physics -- making these experiments indispensable tools in the search for new fundamental interactions.

\section*{Acknowledgements} 
\label{sec:acknowledgements}
AK and RKJ would like to acknowledge financial support from the Indo-French Centre for the Promotion of Advanced Research (CEFIPRA) for support of the proposal 6704-4 under the Collaborative Scientific Research Programme. RKJ also acknowledges support from the IISc Research Awards 2024 and SERB, Department of Science and Technology, GoI through the MATRICS grant~MTR/2022/000821. AK thanks Suvodip Mukherjee for the insightful discussion on CMB experiments.
AK and RKJ would also like to thank Joseph Eimer and Keisuke Inomata for highlighting their works. AK and RKJ also thank Basundhara Ghosh, Brijesh Kanodia, Jishnu Sai P, Rathul N Raveendran, Subhadip Bouri and Yashi Tiwari for useful discussions and insightful comments on the work. Finally, we would like to thank the
referee for their insightful comments and suggestions that have helped to improve the
manuscript.

\onecolumngrid
\appendix

\section{Derivation of the conversion probability}
\label{appendix}
Let us start with the equation which describes the graviton-photon oscillation for the channel $(\tilde{h}_L \leftrightarrow A_+)$~\cite{2024-Kushwaha.Jain-PRD}
\begin{align}\label{appeq-oscillation-eq-matrix}
	\left[ (\omega + i\partial_z) \mathbb{I} + \mathcal{M} \right]   \hat{\Psi} (z)  \simeq 0
\end{align}
where $\mathbb{I}$ is the identity matrix, $\mathcal{M}$ is analogous to mixing matrix and $\hat{\Psi} (z)$ is the column vector field, given by
\begin{align}\label{appeq-mixing-matrix-def}
  \mathbb{I} = \begin{bmatrix}
    1 & 0 \\ 0 & 1
\end{bmatrix}, \quad  \mathcal{M} = \begin{bmatrix}
        -\frac{\omega_{Pl}^2 }{2\omega} & -\frac{\bar{B}}{\sqrt{2} M_P} \\
        -\frac{\bar{B}}{\sqrt{2} M_P} & 0
    \end{bmatrix},    \quad \text{and} \quad
      \hat{\Psi} (z) = \begin{bmatrix}
    A_+ (z) \\ \tilde{h}_L (z)
    \end{bmatrix} ~ ~ .
\end{align}
To solve Eq.\eqref{appeq-oscillation-eq-matrix}, we use the Unitary matrix satisfying $\textbf{U} \textbf{U}^T = \textbf{U}^T \textbf{U} = \mathbb{I}$, which diagonalizes the mixing matrix $\mathcal{M}$ as $\text{diagonal} (\lambda_1, \lambda_2)$ where $\lambda_{1,2}$ are the eigenvalues of $\mathcal{M}$, which are given by
\begin{align}\label{lambda-1-2}
    \lambda_{1,2} = - \frac{1}{2} \left[ \frac{\omega_{Pl}^2 }{2\omega} \pm \sqrt{\left( \frac{\omega_{Pl}^2}{2\omega} \right)^2 + \frac{2 {\bar{B}}^2}{M_P^2}} \right] ~~.
\end{align}
%
%
%
%
The solutions of the Eq.\eqref{appeq-oscillation-eq-matrix} can be given as
\begin{align}\label{psi-equation}
  \Psi (z) =  
  \exp \left\{i \int_0^z (\textbf{U}^T \tilde{\mathcal{M}} \textbf{U} ) dz \right\} \cdot \exp \left\{i \int_0^z \omega dz \right\} \hat{\Psi}(0)
\end{align}
where $\tilde{\mathcal{M} } = \mathbf{U} \mathcal{M} \mathbf{U}^T$ and $e^{i \int_0^z \omega dz }$ is a phase factor and can be absorbed in $\hat{\Psi} (0)$. 
Following our work~\cite{2024-Kushwaha.Jain-PRD}, the solution of the graviton-photon coupled equation \eqref{appeq-oscillation-eq-matrix} (i.e., for $\tilde{h}_L \longleftrightarrow A_+$) can be given by
\begin{subequations}\label{photon-graviton-solution-main-L-plus}
	\begin{alignat}{2}
		    A_+ (z) &\simeq \left( \cos^2 \theta ~ e^{i \lambda_1 |z|} + \sin^2 \theta ~ e^{i \lambda_2 |z|} \right) A_+ (0) + \left( \left( e^{ i\lambda_1 |z|} - e^{i \lambda_2 |z|} \right) \sin\theta \cos\theta \right) \tilde{h}_L (0)
		\\
		\tilde{h}_L (z) &\simeq  \left( \left( e^{ i\lambda_1 |z|} - e^{i \lambda_2 |z|} \right) \sin\theta \cos\theta \right) A_+ (0) +  \left( \sin^2 \theta ~ e^{i \lambda_1 |z|} + \cos^2 \theta ~ e^{i \lambda_2 |z|} \right) \tilde{h}_L (0) ~~.
	\end{alignat}
\end{subequations}
and similarly for $\tilde{h}_R \longleftrightarrow A_-$:
\begin{subequations}\label{photon-graviton-solution-main-R-minus}
	\begin{alignat}{2}
		    A_- (z) &\simeq \left( \cos^2 \theta ~ e^{i \lambda_1 |z|} + \sin^2 \theta ~ e^{i \lambda_2 |z|} \right) A_- (0) + \left( \left( e^{ i\lambda_1 |z|} - e^{i \lambda_2 |z|} \right) \sin\theta \cos\theta \right) \tilde{h}_R (0)
		\\
		\tilde{h}_R (z) &\simeq  \left( \left( e^{ i\lambda_1 |z|} - e^{i \lambda_2 |z|} \right) \sin\theta \cos\theta \right) A_- (0) +  \left( \sin^2 \theta ~ e^{i \lambda_1 |z|} + \cos^2 \theta ~ e^{i \lambda_2 |z|} \right) \tilde{h}_R (0) ~~.
	\end{alignat}
\end{subequations}
Note that for a given set of initial conditions, the above solutions determine the amplitude of the conversion from GW to EM sector and vice versa. Since we are interested in the graviton to photon conversion, we assume that there are only gravitons at the beginning and \emph{no} photons. This leads to the following initial conditions: $\tilde{h}_{L,R} (0) = 1$ and $A_{+,-} (0) = 0$, and using them, we obtain the following results
\begin{subequations}\label{R-minus-L-plus-eqs}
	\begin{alignat}{2}
          \langle \tilde{h}_L (0) | A_+ (z) \rangle &\simeq \left( e^{ i\lambda_1 |z|} - e^{i \lambda_2 |z|} \right) \sin\theta \cos\theta \langle \tilde{h}_L (0) | \tilde{h}_L (0) \rangle ~ , 
            \\
            \langle \tilde{h}_R (0) | A_- (z) \rangle &\simeq \left( e^{ i\lambda_1 |z|} - e^{i \lambda_2 |z|} \right) \sin\theta \cos\theta  \langle \tilde{h}_R (0) | \tilde{h}_R (0) \rangle 
             \simeq \frac{1}{\delta_g} \left( e^{ i\lambda_1 |z|} - e^{i \lambda_2 |z|} \right) \sin\theta \cos\theta  \langle \tilde{h}_L (0) | \tilde{h}_L (0) \rangle
	\end{alignat}
\end{subequations}
where we have used the relation $|h_L | = \sqrt{\delta_g} |h_R|$ and $ \langle \tilde{h}_L (0) | \tilde{h}_L (0) \rangle $ representing the overall normalization term, which can be set to unity. 
To compare the conversion amplitudes in both channels consistently, we must apply the same normalization. This introduces an asymmetry between the two channels governed by $\delta_g$, which quantifies the imbalance between the left and right circular polarization modes of the GWs and is related to the chirality parameter in the GW sector
\begin{align}\label{L-R-conv-prob-relation-term}
     \langle \tilde{h}_R (0) | A_- (z) \rangle = \frac{1}{\delta_g}  \langle \tilde{h}_L (0) | A_+ (z) \rangle ~~.
\end{align} 
\\
Let us first consider the channel $\tilde{h}_L \longleftrightarrow A_+$, and compute the conversion probability, which allows us to calculate the other channel straightforwardly. Using Eq.~\eqref{R-minus-L-plus-eqs}, we have
\begin{align}\label{conv-probability-1}
    \left| \langle \tilde{h}_L (0) | A_+ (z)  \rangle \right|^2 = \sin^2\theta \cos^2\theta  \left| e^{ i\lambda_1 |z|} - e^{i \lambda_2 |z|} \right|^2
    = \sin^2 (2\theta) ~ \sin^2\left( \frac{(\lambda_1 - \lambda_2) |z|}{2} \right)
\end{align}
where the mixing angle $\theta$ is the ratio of the off-diagonal term to the difference of the diagonal terms in the mixing matrix $\mathcal{M}$, and is given by $\tan(2\theta) = \frac{2\sqrt{2} \omega \bar{B} }{\omega_{Pl}^2 M_P} $~\cite{1988-Raffelt.Stodolsky-PRD}. Therefore, the probability of conversion of left-circular gravitons to right-circular photons can be estimated by
using Eq.\eqref{lambda-1-2} in Eq.\eqref{conv-probability-1} as
\begin{align}\label{conv-probability-2}
 \mathcal{P}^+_{g \rightarrow \gamma} (\ell) =    \left| \langle \tilde{h}_L (0) | A_+ (z)  \rangle \right|^2
    =\left( \frac{8\omega^2 {\bar{B}}^2}{\omega_{Pl}^4 M_P^2 + 8 \omega^2 {\bar{B}}^2} \right) ~ \sin^2\left[ \frac{|z|}{2} \sqrt{ \left( \frac{\omega_{Pl}^2}{2\omega} \right)^2 + \frac{2 {\bar{B}}^2}{M_P^2} } \right] ~~.
\end{align}
where $|z| = \ell$ is the typical length scale over which the conversion of GWs to EM waves happens. The analysis up to this point is generic in the sense that, the above expression gives the conversion probability of gravitons to photons as long as the refractive index of the medium is dominated by the plasma frequency $\omega_{Pl}^2$ (other subdominant contributions could be non-linear correction to electrodynamics and Cotton-Mouton effect~\cite{2021-Domcke.Garcia-Cely-PRL,2024-He.Sharma.etal-JCAP}). 
To study the phenomenological consequences of the graviton-photon oscillation for the above equation during the CMB epoch, we define the quantity (i.e., the oscillation length).
\begin{align}\label{oscillation-length}
	\ell_{osc}^{-1} =  \frac{1}{2} \sqrt{ \left( \frac{\omega_{Pl}^2}{2\omega} \right)^2 + \frac{2 {\bar{B}}^2}{M_P^2} }  = \frac{1}{2} \sqrt{ \omega^2 (1-n)^2 + \frac{2 {\bar{B}}^2}{M_P^2} } ~,
\end{align}
and the overall amplitude 
can be simplified as
\begin{align}
	\frac{8\omega^2 {\bar{B}}^2}{\omega_{Pl}^4 M_P^2 + 8 \omega^2 {\bar{B}}^2}  = \frac{2}{M_P^2} \left[ \frac{ {\bar{B}}^2}{ \left( \frac{\omega_{Pl}^2}{2\omega} \right)^2  +  \frac{2\bar{B}^2}{M_P^2} } \right] = \frac{\bar{B}^2 \ell^2_{osc}}{2 M_P^2} ~~.
\end{align}
Substituting the above simplifications in Eq.\eqref{conv-probability-2} gives the conversion probability as
\begin{align}\label{conv-probability-2app}
 \mathcal{P}^+_{g \rightarrow \gamma} (\ell) =   \frac{\bar{B}^2 \ell^2_{osc}}{2 M_P^2} ~ \sin^2\left[ \frac{\ell}{\ell_{osc}}\right] ~~.
\end{align}
Note that the relation \eqref{L-R-conv-prob-relation-term} allows us to straightforwardly compute the conversion probability for the other channel i.e., $\mathcal{P}^-_{g \rightarrow \gamma} (\ell) = \frac{1}{\delta_g^2}\mathcal{P}^+_{g \rightarrow \gamma} (\ell)$.

\bibliographystyle{apsrev4-1}
\input{References.bbl}
\end{document}

%% file: References.bbl
%